\documentstyle[12pt,moriond]{article}
\begin{document}

\input epsf  

% produces <~ or >~ signs
\def\spose#1{\hbox to 0pt{#1\hss}}
\def\simlt{\mathrel{\spose{\lower 3pt\hbox{$\mathchar"218$}}
     \raise 2.0pt\hbox{$\mathchar"13C$}}}
\def\simgt{\mathrel{\spose{\lower 3pt\hbox{$\mathchar"218$}}
     \raise 2.0pt\hbox{$\mathchar"13E$}}}
\def\deg{^\circ} 
\def\msun{\thinspace\hbox{$\hbox{M}_{\odot}$}}
\def\reference#1#2#3#4#5{\tenp\par\noindent\hangindent 3em
              #1, #2. {\tenpsl #3\/}, {\tenpb #4,}
\thinspace\hbox{#5}}
\def\etal   {et\nobreak\ al.\ }
\def \aa#1#2   {{\em Astr. Astrophys. \/} {\bf #1}, {#2}}
\def \aas#1#2  {{\em Astr. Astrophys. Suppl. Ser. \/} {\bf #1}, {#2}}
\def \aj#1#2   {{\em Astron. J. \/} {\bf #1}, {#2}}
\def \apj#1#2  {{\em Astrophys. J. \/} {\bf #1}, {#2}}
\def \apjl#1#2  {{\em Astrophys. J. Lett.\/} {\bf #1}, {#2}}
\def \apjs#1#2 {{\em Astrophys. J. Suppl. Ser. \/} {\bf #1}, {#2}}
\def \mnras#1#2{{\em MNRAS \/} {\bf #1}, {#2}}
\def \nat#1#2  {{\em Nature \/} {\bf #1}, {#2}}

\heading{THE LOCAL ENVIRONMENT OF HII GALAXIES}
 
\author{Eduardo Telles $^{1}$, Steve Maddox $^{2}$} {$^{1}$ Observatorio Nacional, Rio de Janeiro,              Brasil.}  {$^{2}$ 
Institute of Astronomy,  Cambridge, U.K.}

%\begin{figure*}[h]
%\protect\centerline{
%\epsfxsize=3.2cm\epsffile{me2.ps}
%}
%\end{figure*}
 
\vspace{10mm} 
 
\begin{moriondabstract}

We address the question of whether violent star formation in HII
galaxies is induced by low mass companions by describing statistically
their local environment as estimated by the correlation function.  We
argue that even if low mass companions were mainly intergalactic HI
clouds, their optical counterparts should be detectable at faint
limits of the Automatic Plate Measuring Machine scans.  We then
cross-correlate a large sample of HII galaxies with the APM faint
field galaxy catalogue.  The preliminary results are all consistent
with HII galaxies being a randomly selected sample of normal faint
field galaxy with no extra clustering.  This suggests that at least in
these dwarf starburst galaxies star formation is not triggered by
tidal interactions and may have a different origin.

\end{moriondabstract}

\vspace{10mm} 
 
\section{Introduction}

HII galaxies are dwarf galaxies in a bursting phase of star formation
of low luminosity (mass), low heavy element abundance and low dust
content where the triggering mechanism of the present episode of
violent star formation is not so obvious \cite{tt}.

% Their optical properties are dominated by the massive star forming
% region as shown by their strong emission line spectra superposed on a
% weak blue continuum.  The properties of the underlying galaxies in
% these systems are similar to late type dwarf galaxies such as dwarf
% irregulars or low surface brightness dwarfs (Telles \& Terlevich
% 1997).  The most luminous HII galaxies, classified as Type I's by
% Telles, Melnick and Terlevich (1997), show signs of disturbed
% morphology such as distorted outer isophotes, tails or irregular fuzz,
% while the low luminosity Type II's are regular and compact.  Although
% there is a clear case for a morphology-luminosity relation, neither
% type of HII galaxies shows conspicuous evidence of bright companions
% in their neighbourhood.  The few HII galaxies found to have a bright
% neighbour (maybe by chance) are all of Type II's of regular
% morphology, contrary to what one would expect if interactions caused
% the morphological disturbances as seen in Type I's (Telles \&
% Terlevich 1995).
% 
% A popular hypothesis is that interactions between galaxies are the
% triggers of starbursts and they may also cause the current burst of
% star formation in HII galaxies.  

HII galaxies are less clustered than normal bright galaxies and tend
to populate regions of low galactic density
\cite{ims}\cite{cm}\cite{cmm}\cite{pults}\cite{rsm}\cite{tt}\cite{v}.
They are not associated with giant galaxies, therefore, HII galaxies
are not tidal debris of strongly interacting systems.

% A possible alternative was presented by Melnick (1987) proposed that
% high resolution 21cm maps were needed to investigate the role of
% collisions between intergalactic neutral hydrogen clouds in the
% formation of these objects.  Brinks (1990) also hypothesized that
% other dwarfs or intergalactic HI clouds could be the triggering
% agents.  Taylor \etal (1995, 1996) using the VLA detected12 HI
% companions around 21 HII galaxies, while only 4 HI-rich companions
% were detected around a control sample of 17 low surface brightness
% dwarfs (Taylor 1997).  

A possible and appealing alternative that other dwarfs or
intergalactic HI clouds could be the triggerers \cite{b}\cite{m} was
followed up by Taylor and collaborators
\cite{tbgs1}\cite{tbgs2}\cite{ta}.  They have used the VLA 21 cm maps
to search for HI companions around HII galaxies.  The main VLA results
can be summarized as follows: (i) 12 out of 21 HII galaxies have HI
companions; (ii) 13 out of 17 Low Surface Brightness Galaxies (LSBG)
from a control sample of quiescent galaxies do not have HI
companions. As also pointed out by these authors, some questions
remain intriguing from this: Why are these 9 out of the 21 HII
galaxies with \underline{no} companions violent forming stars now
('bursting')?  Why are these 4 out of the 17 LSBGs with companions
{\em not} 'bursting'?

Since most of the detected clouds in HI surveys have optical
counterparts at faint levels (c.f. Hoffman and Zwaan at this meeting),
we have carried out a further investigation of the galaxy environments
of a unbiased sample of over 160 low redshift HII galaxies by
cross-correlating their accurate position in the sky, also derived
from the APM scans, to faint field galaxies ($15 < b_{\rm J} < 20$)
galaxies in the APM galaxy catalogue.  The HII galaxy sample used in
this work is taken from the {\em Spectrophotometric Catalogue of HII
Galaxies} \cite{tmmmc}. Our sample of faint field galaxies was
selected from the APM Galaxy Survey, which is described in detail in
\cite{msegl}. The galaxy sample selected from the survey data at a
magnitude limit of $b_{\rm J} =20.5$ has a completeness
$\sim90$--$95\%$, stellar contamination $\sim 5\%$ \cite{mes}.

% Most, if not all, of these HI companions have
% optical counterparts at faint levels.  Taylor \etal also estimate
% these HI companions to have HI masses larger than a few $10^7$ \msun.
% If we consider a HI mass - luminosity relation for dwarf galaxies, for
% instance one derived from the observations of Thuan \& Martin (1981),
% we can expect these companions to have absolute magnitudes typical of
% dwarf galaxies (M$_{B} \leq -15$).  At the range of redshift of our
% HII galaxy sample, these companions should be detectable at the very
% faint optical levels of the Automatic Plate Measuring Machine (APM)
% scans of UK Schmidt plates.

\section{ The angular cross-correlation function $w_{hg}(\theta)$ }

\begin{figure*}
\protect\centerline{
\epsfxsize=3in\epsffile[ 60 195 520 645]{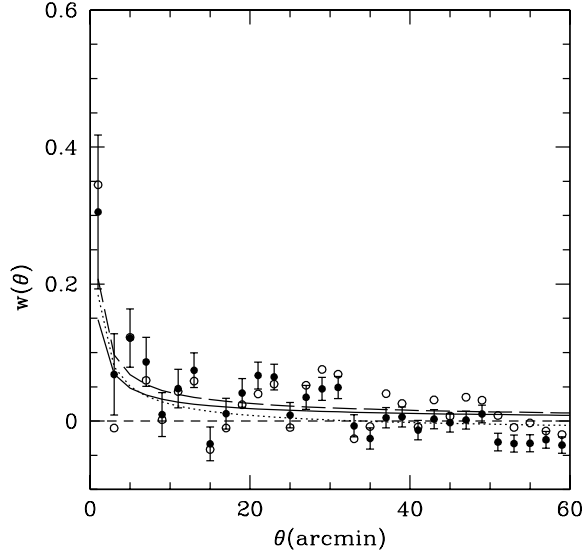} 
}
\caption{\label{wtheta}{\small The angular cross-correlation between
the HII galaxies and the faint APM field galaxies.}}
\end{figure*}

The angular cross-correlation function $w_{hg}(\theta)$ is estimated
by comparing the number of galaxies $N_{HG}$ as a function of angular
radius $\theta$ from the central HII galaxy with the number $N_{HR}$
counted for a catalogue of uniform random positions with the same
surface density:
\begin{equation} 
w_{hg}(\theta) = {{ N_{HG}(\theta)} \over {N_{HR}(\theta)} }   -1
\label{whg} 
\end{equation} 
We also used the simpler direct estimate using the mean surface
density of field galaxies.
This gave essentially indistinguishable results.

The main results are shown in Figure~\ref{wtheta}.  The filled points
show $w$ from equation~\ref{whg} and the open points from the direct
estimator.  It can be seen that $w_{hg}$ is significantly positive for
angles $\theta\simlt 10 '$, and this corresponds to an excess of
galaxies near the HII galaxy positions over a uniform distribution.
The lines in Figure~\ref{wtheta} show the predicted cross-correlation
between the faint APM sample and a field galaxy sample with the
same redshift distribution as the HII galaxy sample.  To predict this,
we used the measured APM $w(\theta)$ \cite{mes}, and calculated
scaling factors by numerically integrating Limbers equation \cite{p}
with the HII galaxy redshift distribution and the 
APM redshift distribution as given by \cite{mes}.
Figure~\ref{wtheta} shows that the estimated $w_{hg}(\theta)$ for HII
galaxies (points) follows the prediction (lines) for what we would
expect if HII galaxies were clustered in the same way as normal faint
field galaxies.  
There is a marginal excess over the prediction, but this is not
significant compared to the expected errors, and  we conclude that
HII galaxies have the same number of companions as faint
field galaxies.

%calculation 
%results - a picture 
%comparison to expected for normal galaxies

\section{ The projected cross-correlation function  %$\xi_{hg}$ }
$\Xi_{hg}(\sigma)$ }

\begin{figure*}
\protect\centerline{
\epsfxsize=3in\epsffile[ 60 195 520 645]{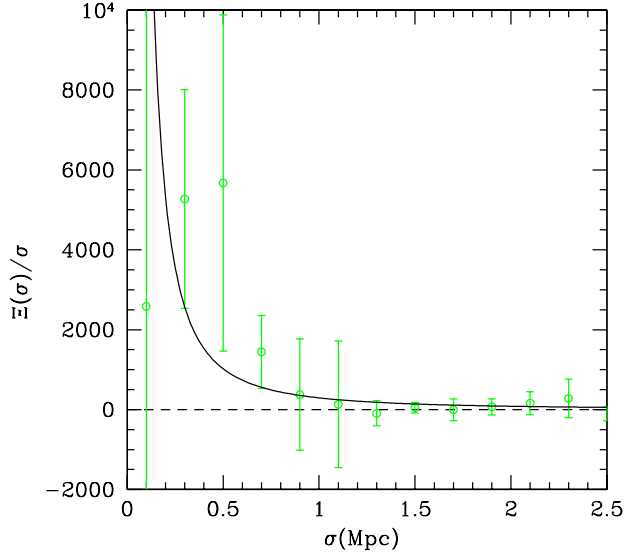}
}
\caption{\label{wsigmar}
{\small The observed projected cross-correlation between the HII galaxies and
the faint APM field galaxies, $\Xi_{hg}(\sigma)/\sigma$. 
}}
\end{figure*}

Since we know the redshift to each HII galaxy, we can estimate the
correlation function using physical separation in kpc.  The resulting
projected cross-correlation function is an integral over the spatial
correlation function $\xi_{hg}$, and traditionally denoted by
$\Xi_{hg}(\sigma)$, where $\sigma$ is the projected separation.  For a
simple power law correlation function, 
$\Xi_{hg}(\sigma)/\sigma = \xi_{hg}(\sigma) \Gamma(
\frac{1}{2} ) \Gamma( \frac{\gamma-1}{2} ) \Gamma( \frac{\gamma}{2} )$.
We defer the formal analysis to a forthcoming paper and present our
preliminary results in Figure~\ref{wsigmar}.

It can be seen that the measured $\Xi_{hg}$ is positive for $\sigma
\simlt 1Mpc$.  We have also calculated the expected clustering,
$\Xi_{hg}(\sigma)/\sigma$, on the assumption that HII galaxies are
clustered in the same way as normal galaxies on small scales,
$\xi_{hg}(r) = (r/5.7h^{-1})^{-1.8} $  \cite{mes}.  Note that
the different distance to each HII galaxy means that relation between
$\sigma$ and $\theta$ is different for each centre, and also the $
1/n(x) $ leads to a different weighting of the pair count from each
centre.  This means that $\Xi_{hg}$ is not simply a rescaling of
$w_{hg}$.  As in the case of the angular cross-correlation, the
measured value here is consistent with the prediction. 

%The local density of galaxies around a typical HII galaxy. 
%The result is shown in figure~x 

%calculation - direct ensemble estimators. 
%results - a picture 
%comparison to expected for normal galaxies

Assuming that all of the excess galaxies compared to a random
distribution, $\sigma < 1$ Mpc, are at the redshift of the central
galaxy, the magnitude distribution of the excess directly gives the
luminosity function (LF) of the neighbouring galaxies \cite{srl}.
Comparison with the LF of faint field galaxies estimated from the
Stromlo/APM survey \cite{lpem}\cite{l} shows that the galaxies near
to HII galaxies have an absolute magnitude distribution consistent
with normal galaxies.

\section{Conclusion}

\begin{enumerate}

\item Both the angular and projected correlation functions are
significantly above zero, so HII galaxies are significantly clustered.
This is what you expect to find for any sample of galaxies. 

\item Both the angular and projected measurements are consistent with
the predictions expected for a sample of normally clustered galaxies,
showing that HII galaxies are no more or less clustered than normal
{\em faint} field galaxies.

\item The LF from the extra galaxies within 1 Mpc compared to a random
distribution shows that the galaxies near to HII galaxies have an
absolute magnitude distribution consistent with normal faint field
galaxies.   This result will be shown in a forthcoming paper.

\end{enumerate}

In summary, HII are less clustered than bright galaxies, but our
present results are all consistent with HII galaxies being a randomly
selected sample of normal {\em faint} field galaxies.  This suggests
that at least in these dwarf galaxies (or in most of them) star formation
is not triggered by tidal interactions and may have a different
origin.

\vfill
\end{document}